\documentclass[useAMS,usenatbib,usegraphicx]{mn2e}

\newcommand{\kev}{keV}

\newcommand{\fe}{Fe~K$\alpha$}
\newcommand{\etal}{et al.}
\newcommand{\threec}{3C~120}

\title[The \textit{XMM-Newton} view of 3C~120]
  {The \textit{XMM-Newton} view of the broad-line radio galaxy 3C~120}
\author[D.\ R.\ Ballantyne \etal]
  {D.~R.~Ballantyne,$^1$\thanks{ballantyne@cita.utoronto.ca}
  A.~C.~Fabian,$^2$ and K.~Iwasawa$^2$\\
  $^1$Canadian Institute for Theoretical Astrophysics, 60 St. George
  Street, Toronto, Ontario, Canada M5S 3H8 \\
  $^2$Institute of Astronomy, Madingley Road, Cambridge CB3 0HA}

\pagerange{\pageref{firstpage}--\pageref{lastpage}}
\pubyear{2004}

\usepackage{times}

\begin{document}

\label{firstpage}

\maketitle

\begin{abstract}
We present the results of a 127~ks \textit{XMM-Newton} observation of
the broad-line radio galaxy \threec\ performed simultaneously with
\textit{RXTE}. The data have yielded the highest quality 0.6--10~\kev\
spectrum of a radio-loud active galaxy ever produced. The
time-averaged spectrum is Seyfert-like, with a reflection amplitude
$R\sim 0.5$, and a neutral \fe\ line with equivalent width $\sim
53$~eV. The line is slightly broadened with a FWHM$\approx
10^4$~km~s$^{-1}$. This is consistent with arising from an accretion
disc radius of $\ga 75$~$GM/c^2$ at an inclination angle of $\sim
10$~degrees, consistent with the limit of $< 14$ degrees derived from
the radio jets. At low energies the spectrum requires excess
absorption above the Galactic value and a soft excess which is best
fit with a bremsstrahlung model ($kT=0.3$--$0.4$~keV). The total
luminosity in the bremsstrahlung component is just under half of the total
hard X-ray luminosity. The emission may originate in either the broad
line region, or in giant H~\textsc{ii} regions adjacent to the
nucleus. Weak O~\textsc{vii} and O~\textsc{viii} edges are detected
with high precision, suggesting the presence of a warm absorber
component. Broadband 0.6--50~\kev\ fits to the data cannot
unambiguously determine the accretion mode in \threec. A two-component
ionized disc model, with a very highly ionized reflector presumably
arising from very close to the black hole, is only a small improvement
over a truncated disc model. The strength of the soft X-ray emission
features produced by the distant neutral reflector are overpredicted
in our solar abundance model, implying that the heavy metal abundance
in \threec\ is subsolar. Both broadband models could also fit a much
shorter archival \textit{XMM-Newton} observation. The total count-rate
declined by 20 per cent over the course of the long observation, while
small-scale rapid variability was present at the level of a few
percent.  A possible increase in the \fe\ line flux, significant
at the 90 per cent level, was identified at $\sim 80$~ks in the
observation. The total unabsorbed luminosity of \threec\ implies that
it is accreting close to its Eddington rate, consistent with a model
of a highly ionized thick disc. A possible connection between
accretion disc thickness and radio jet production is discussed.
\end{abstract}


\section{Introduction}
\label{sect:intro}
\threec\ ($z=0.033$; \citealt{burb67}) is the brightest broad-line
radio galaxy (BLRG) in the X-ray sky ($F_{\mathrm{2-10\ keV}} =
4.2\times 10^{-11}$~erg~cm$^{-2}$~s$^{-1}$; \citealt*{sem99}). It has
been a well known and well studied source at almost every wavelength
since the discovery of its optical counterpart \citep{cbs66}. While
the optical spectrum of \threec\ is very typical of a Seyfert 1, it
resides in an optically peculiar galaxy that shows only some
indication of spiral structure \citep{arp75,mol88}. The galaxy has a very
random and confused velocity field with only slight evidence of
rotation \citep{bald80,mol88}. Photoionized gas is common in the
galaxy, and was interpreted by \citet{bald80} as resulting from
illumination by the nuclear continuum. However, more recent spectra of
the nebulosity showed that the gas was at least partially photoionized
by a significant episode of star formation in the galaxy
\citep{mol88,sou89}. This fact, together with the unusual velocity
field, has lead to the hypothesis that \threec\ is in the late stages
of a merger (e.g., \citealt{heck86}; although see comments in
\citealt{bald80}). The optical continuum of the nucleus is known to be
variable \citep[e.g.,][]{wwg79,webb90}, as are the emission lines
\citep{ors80,fm80,pet98}, thus allowing reverberation
mapping to yield a well-constrained black hole mass of
$3.0^{+2.0}_{-1.5}\times 10^7$~M$_{\odot}$ \citep{wpm99}.

At radio wavelengths, \threec\ is core-dominated and exhibits a
superluminal jet ranging from sub-pc to nearly 100 kpc scales
\citep*{coh77,sei79,wbu87,walk01}. The superluminal speed results in
an upperlimit to the inclination angle of the jet --- and, presumably,
the accretion disc --- to the line of sight of just 14 degrees
\citep{eh98}. Ejections of plasma into the jet may be connected to
dips in the X-ray flux \citep{mar02}.

The X-ray emission from the nucleus of \threec\ has been observed
since the beginning of satellite-based X-ray astronomy and was known
to be a variable source with a canonical power-law spectral shape that
softened as the source brightened (\citet{mar91} and references
therein). A weak correlation between the highest frequency UV emission
detected by \textit{IUE} and the softest X-ray emission from
\textit{EXOSAT} \citep{mar91} indicated a Comptonization origin for
the X-rays, rather than synchrotron emission from the jet. \threec\
was not observed by \textit{Ginga}, so the first detection of \fe\
emission was made by \textit{ASCA}. This 50~ks observation clearly
detected an \fe\ line, but despite five different analyses of this
dataset there has been no consensus on the strength of the
line. Workers who modeled the continuum as an absorbed power-law
\citep{rey97,sem99} consistently found the line to
be very broad ($\sigma > 1.5$~keV) and very strong (equivalent width
[EW] $\sim 1000$~eV). If, however, the continuum was modeled with
reflection \citep{gr97}, or with a broken power-law
\citep{woz98}, the line width and EW dropped by over a
factor of two. This is consistent with the results of \citet{nan97},
who fit a power-law to the 3--10~\kev\ data. More recent
observations of \threec\ by \textit{RXTE} \citep*{esm00,gse03} and
\textit{BeppoSAX} \citep{zg01} have provided some
consistency by requiring reflection to fit the continuum and obtained
an Fe~K$\alpha$ EW $\sim 100$~eV. Thus, previous work has shown that
\threec\ has a curved X-ray continuum with spectral hardening at high
energies and a Fe~K$\alpha$ line which may be broad.

A common result from the reflection models fit to these data is a weak
reflection normalization (i.e., $\Omega/2\pi=0.4$;
\citealt{gse03}). This fact, together with the $\sim 100$~eV iron
K$\alpha$ EW, led \citet{esm00} and \citet{zg01} to postulate that the
optically thick accretion disc is truncated in \threec\ to a hot,
optically thin flow. Indeed, based on similar evidence, this
interpretation has been applied to other BLRGs
\citep{eh98,gr99,esm00}. If true, this change in accretion flow may
provide an important clue to the cause of the radio-loud/radio-quiet
dichotomy in AGN physics. However, other explanations, such as an
ionized accretion disc \citep*{brf02}, can fit the \textit{ASCA} data
of \threec\ without resorting to a change in accretion
geometry. Better data are required before a detailed comparison can be
made between a BLRG and the radio-quiet Seyfert~1s.

This paper presents such data in the form of a simultaneous 127~ks
\textit{XMM-Newton} and \textit{RXTE} observation of \threec. The
resulting 0.6--10~\kev\ spectrum has the highest signal-to-noise of
any BLRG spectrum in that energy range. The observational details are
outlined in the next section, and the data reduction is described in
Section~\ref{sect:data}. The spectral and timing analyses of this
dataset are presented in Sections~\ref{sect:spectral}
and~\ref{sect:timing}, respectively. The results are discussed in
Section~\ref{sect:discuss}, before we end by summarizing in
Section~\ref{sect:concl}.

\section{Observations}
\label{sect:obs} 
\textit{XMM-Newton} \citep{jan01} observed \threec\ during revolution
680 between 2003 August 26 05:12:32 and 2003 August 27 18:50:52. The
telescope contains two major instruments: the European Photon Imaging
Camera (EPIC), which consists of two MOS \citep{tur01} and one pn
\citep{stru01} detector, and the Reflection Grating Spectrometers
(RGS; \citealt{dh01}). This paper focuses on the broadband continuum
of \threec, so we will defer the analysis of the RGS spectra to a
later publication. The medium optical filter was applied to the three
EPIC cameras, and the MOS-2 and pn detectors were operated in small
window mode. The MOS-1 detector was operated in timing mode, and those
data were not included in the following analysis. Finally, the Optical
Monitor (OM; \citealt{mas01}) onboard \textit{XMM-Newton} obtained 26
exposures of \threec\ in the UVW1 band (2450--3200~\AA). Those data
will also be presented in another publication.

\textit{RXTE} observed \threec\ between 2003 August 26 02:55:12 and
2003 August 27 18:59:44, entirely covering the duration of the
\textit{XMM-Newton} observation. The telescope contains two collimated
instruments that together cover the energy range 3--250~\kev. The
Proportional Counter Array (PCA; \citealt{jah96}) consists of 5 xenon
and propane-filled co-aligned proportional counter units (PCUs). Only
PCU0 and PCU2 were switched on for the duration of the \threec\
observation, with the other three units taking data
intermittently. The High Energy X-ray Timing Experiment (HEXTE;
\citealt{rot98}) contains two clusters each with four phoswich
detectors. During an observation, the clusters are rocked on and off
source so that one is collecting data from the source while the
other is observing the background.

\section{Data Reduction}
\label{sect:data}

\subsection{\textit{XMM-Newton}}
\label{sub:xmm}
The EPIC data were extracted from the Observation Data Files (ODF)
using the 'emchain' and 'epchain' tasks within the \textit{XMM-Newton}
Science Analysis System (SAS) v.6.0. These chains generated calibrated
event lists after removing bad pixels, and applying both gain and
Charge Transfer Inefficiency (CTI) corrections\footnote{see, e.g.,
http://xmm.vilspa.esa.es/sas/current/doc/epchain/index.html}.
\threec\ fell near the edge of the pn small window, and so the
spectrum was extracted from a circular region with a radius of 45
arcseconds.  The background region was also defined by a circle with
the same radius in one of the corners of the window. The background
was relatively steady during the observation except for a large
increase in count rate during the last 4--5~ks of the observation when
the telescope was approaching perigee. This period was not included in
the data extraction. A circular region with a radius of 25 arcseconds
was employed to extract data from the MOS-2 small window. A
similar circle on an adjacent CCD was used to obtain background
data for this detector. With these extraction procedures, we obtained
a pn dataset with a total good exposure time of 87.5~ks (including the
$\sim 71$ per cent live-time of the pn in small window mode;
\citealt{stru01}) which contained $1.8\times 10^6$~photons when both single and double
events were included. The MOS-2 detector yielded $6.8\times 10^5$
photon in 123~ks of good exposure time. Single, double, triple and
quadruple events were included in those data. Response matrices and
ancillary response files for all the datasets were generated with the
'rmfgen' and 'arfgen' tools.

The SAS task 'epatplot' was used to determine if the extracted spectra
were affected by pileup \citep{bal99}. Unfortunately, the observed
count rate was high enough (5.5~s$^{-1}$) that the MOS-2 spectra would
be distorted by pileup. Therefore, we do not include the MOS-2 data in
the following spectral analysis. The pn data were unaffected by pileup.

\subsection{\textit{RXTE}}
\label{sub:rxte}
The PCA and HEXTE data were reduced using the tools in FTOOLS v.5.3.,
with the assistance of the 'rex' script. The PCA data were screened
using the standard criteria. Events were accepted if the Earth
elevation angle was $\geq 10$ degrees, the pointing offset was $\leq
0.02$ degrees, the spacecraft had been out of the South Atlantic
Anomaly for at least 30 minutes, and the ELECTRON2 parameter (a
measure of the background in PCU2) was less than 0.1. Data were then
extracted from the top layer of PCU0 and PCU2. The background model
appropriate for faint extragalactic sources,
\texttt{pca\_bkgd\_cmfaintl7\_e5v20030331.mdl}, was used in the
reduction. The tool 'pcarsp' was then used to produce the response
matrix for the time averaged data.

The HEXTE data were also screened by the above criteria, but the
background is extracted directly from the detector (since the HEXTE
clusters alternate observing the source and the background). One of the
detectors in Cluster-1 can no longer produce spectral
information, therefore only data from Cluster-0 is used in the
analysis presented below. The file \texttt{xh97mar20c\_pwa\_64b.rmf} is
used as the response matrix for Cluster-0 data.

\section{Spectral Analysis}
\label{sect:spectral}
In this section we analyze the time-averaged spectrum of \threec\ from
all three instruments (EPIC-pn, PCA, and HEXTE). The spectra were
grouped to have a minimum of 20 counts per bin in order for $\chi^2$
minimization to be applicable. The spectral fitting was performed
simultaneously on all datasets using XSPEC v.11.3.0p
\citep{arn96}. The uncertainties on the best fit parameters are the
2$\sigma$ errorbars for one parameter of interest (i.e., $\Delta
\chi^2=2.71$). Galactic absorption toward \threec\ is taken to be
$N_{\mathrm{H}}=1.20\times 10^{21}$~cm$^{-2}$ \citep{elw89}, and is
modeled by the 'TBabs' code \citep{wam00} in XSPEC. Unless otherwise
specified, all fits restricted the energy range of the various
instruments as follows: 0.6--10~\kev\ (pn), 7--20~\kev\ (PCA), and
20--50~\kev\ (HEXTE). The spectrum hardens suddenly and drastically
($\Delta \Gamma \sim 1.5$) at energies less than 0.6~\kev. It is
unknown if this is real or a calibration problem (the hardening
appears below 0.5~\kev\ in the MOS-2), so we have ignored those data
pending further investigation. To account for any cross-calibration
uncertainties between the different instruments, their relative
normalizations were left free during the spectral fitting process.

\begin{figure}
\centerline{
\includegraphics[width=0.35\textwidth,angle=-90]{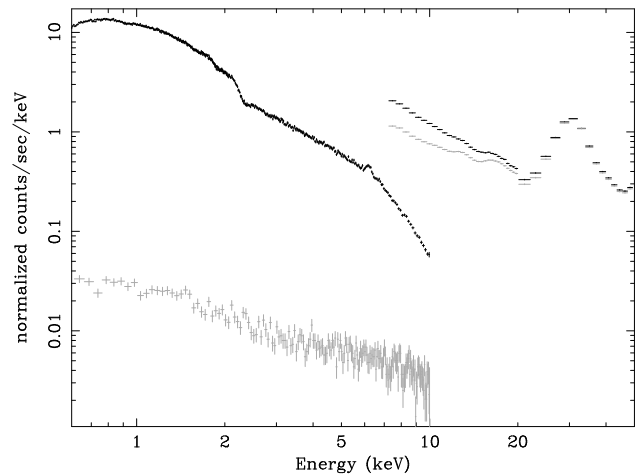}
}
\caption{Observed source (black) and background (grey) count spectra
  for the EPIC-pn (0.6-10~keV), PCA (7--20~keV) and HEXTE (20--50~keV).}
\label{fig:spectra+bkgs}
\end{figure}

In Figure~\ref{fig:spectra+bkgs}, the observed
source and background spectra are plotted for all three instruments
indicating that, except for the HEXTE data, the source is much higher
than the background over the energy range of interest. 

\subsection{The general spectra shape}
\label{sub:general}
\begin{figure}
\centerline{
\includegraphics[width=0.35\textwidth,angle=-90]{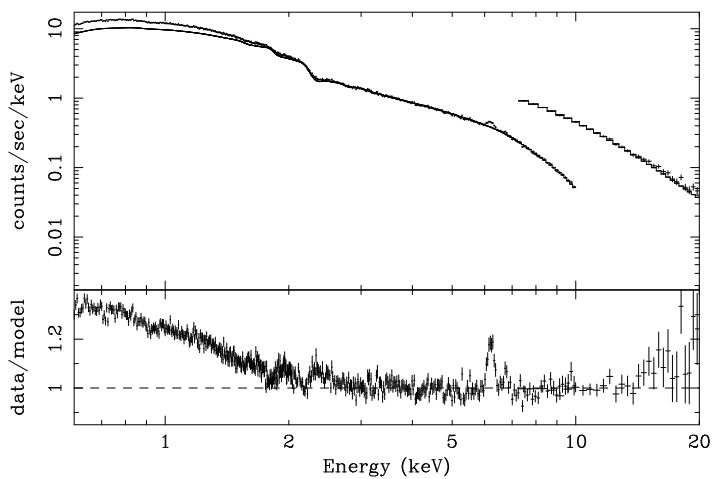}
}
\caption{The upper panel shows the \threec\ count spectra between 0.6
  and 20~\kev\ from the pn and PCA detectors, while the bottom panel
  plots the data-to-model ratio assuming a model of an absorbed
  power-law fit between 3 and 50~\kev. A soft excess, a two component
  Fe K line, and a weak reflection hump are all seen in the
  residuals. The HEXTE data, although not shown here, were used in the
  fit.}
\label{fig:broadband}
\end{figure}

To obtain an overview of the shape of the X-ray spectrum, we fit
the pn, PCA and HEXTE data between 3 and 50~\kev\ with a power-law
model altered only by Galactic absorption. The best fit spectral index
was $\Gamma=1.78\pm 0.01$, but was statistically unacceptable
($\chi^2$/d.o.f.=1658/1435; d.o.f.=degrees of freedom). Extrapolating
the model to low energies gives a view of the spectral shape over
nearly two orders of magnitude in energy (Figure~\ref{fig:broadband}).

The residuals show that the spectrum exhibits significant curvature
particularly below $\sim 3$~\kev\ where a soft excess is required. A
soft excess was not necessary in the \textit{ASCA} analyses
\citep{rey97,gr97} of \threec, but one was found in both the
\textit{ROSAT} \citep{gr97} and \textit{BeppoSAX} data
\citep{zg01}. These last authors attribute the soft excess to an
extended thermal component possibly seen in the \textit{ROSAT} High
Resolution Imager (HRI). However, \citet{mck03} very recently
presented a \textit{Chandra} gratings observation of \threec, and
found that the object was entirely consistent with being a point
source, with only a very weak ($\sim$6 per cent of peak) extension in one
direction. Therefore, it seems unlikely that the soft excess is due to
thermal emission far away from the AGN.

\begin{figure}
\centerline{
\includegraphics[width=0.35\textwidth,angle=-90]{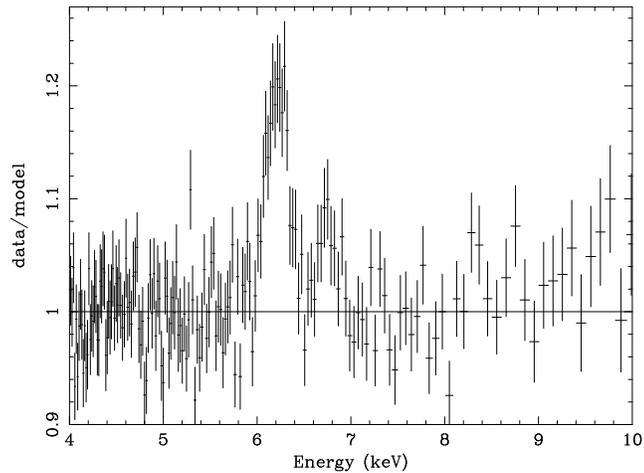}
}
\caption{The data-to-model ratio between 4 and 10~\kev, for an absorbed
  power-law fit to the 3--50~\kev\ data. The data between 5 and
  7.2~\kev\ were ignored in the fit to emphasize the emission features
  in the iron line region.}
\label{fig:feline}
\end{figure}

The hard power-law fit also reveals significant residuals in the \fe\
region (Figure~\ref{fig:feline}), and a weak reflection hump at $\sim
15$--$20$~\kev.
Below, we will first concentrate on describing the hard X-ray
spectrum of \threec, including the \fe\ line and reflection
features. We will then extrapolate these models to the low energy end
to investigate their effect on the soft excess.

\subsection{The hard X-ray continuum and \fe\ line}
\label{sub:hardxrays}
In this section we attempt to describe the spectrum of \threec\
between 3 and 50~\kev\ using both simple and detailed models for the
continuum. The results are summarized in Table~\ref{table:fitres}.
\begin{table*}
\begin{minipage}{175mm}
\caption{Results from fitting the spectrum of \threec\ from 3 to
  50~\kev. The 'Model' column describes the models employed in each
  case with the following notation: PL=power-law, G=Gaussian emission
  line, PEXRAV \& PEXRIV = the reflection continua calculated by
  \citet{mz95}, DL = diskline emission \citep{fab89} and IONDISC =
  constant-density ionized disc reflection model
  \citep*{rf93,rfy99}. $\Gamma$ is the photon-index of the incident
  power-law, $E_{\mathrm{b}}$ is the rest energy of the broad emission
  line (in keV), and $\sigma_{\mathrm{b}}$ is the width of the broad line
  (in keV). The DL parameters tabulated are $r_{\mathrm{in}}$, the
  inner radius of emission in units of $r_g$, $\beta$, the power-law emissivity
  index, and $i$ the inclination angle (in degrees). The reflection
  parameters are $R$, the reflection fraction, and $\xi$, the
  ionization parameter (in erg~cm~s$^{-1}$). When either of the PEXRAV
  or PEXRIV code were in use the inclination angle was fixed at 18.2
  degrees, the minimum allowed by the models.}
\label{table:fitres}
\begin{tabular}{ccccccccc}
Model & $\Gamma$ & $E_{\mathrm{b}}$ & $\sigma_{\mathrm{b}}$ &
$r_{\mathrm{in}}/\beta$ & $i$ & $R$ & $\xi$ &
$\chi^2/$d.o.f.\\ \hline
PL+G & 1.79$\pm 0.01$ & 6.41$\pm 0.02$ & 0.10$^{+0.03}_{-0.02}$ & & & & & 1441/1430 \\
PL+2G$^\dagger$ & 1.80$\pm 0.01$ & 6.41$\pm 0.02$ & 0.10$\pm 0.02$ & & & & & 1419/1228 \\
PEXRAV$^\ddagger$+2G & 1.83$\pm 0.02$ & 6.41$\pm 0.02$ & 0.09$\pm 0.02$ & & 18.2$^f$ & 0.46$^{+0.12}_{-0.11}$ & & 1391/1428 \\
PEXRIV$^\ddagger$+2G & 1.83$\pm 0.02$ & 6.41$\pm 0.02$ & 0.09$\pm 0.02$ & & 18.2$^f$ & 0.42$^{+0.11}_{-0.10}$ & 0.05$^{+1.97}_{-0.05p}$ & 1390/1427 \\
PEXRAV$^\ddagger$+DL$^\ast$+G & 1.83$\pm 0.02$ & 6.44$\pm 0.02$ & & 176$^{+84}_{-49}/-3^f$ & 18.2$^f$ & 0.46$^{+0.11}_{-0.12}$ & & 1389/1428 \\
 & 1.83$\pm 0.02$ & 6.43$^{+0.03}_{-0.02}$ & & 6.0$^f/-0.97^{+1.09}_{-0.49}$ & 18.2$^f$ & 0.47$\pm 0.13$ & & 1394/1428 \\
1xFeIONDISC+G & 1.85$^{+0.01}_{-0.02}$ & & & & & 0.45$^{+0.11}_{-0.06}$ & 52$^{+15}_{-17}$ & 1443/1430 \\
blrd$^\ast$1xFeIONDISC+G & 1.85$^{+0.01}_{-0.02}$ & & & 74$^{+16}_{-25}/-3^f$ & 10.2$\pm 1.5$ & 0.44$\pm 0.04$ & 10$^{+24}_{-0p}$ & 1389/1428 \\
 & 1.86$\pm 0.02$ & & & 6.0$^f/-1.7\pm 0.2$ & 10.4$^{+1.1}_{-10.4p}$ & 0.49$^{+0.06}_{-0.04}$ & 10$^{+21}_{-0p}$ & 1397/1428\\
blrd$^\ast$0.5xFeIONDISC+G & 1.90$\pm 0.02$ & & & 74$^{+16}_{-18}/-3^f$ & 9.5$^{+2.5}_{-3.1}$ & 0.75$^{+0.05}_{-0.09}$ & 12$^{+24}_{-2p}$ & 1389/1428 \\
blrd$^\ast$2xFeIONDISC+G & 1.82$\pm 0.01$ & & & 73$^{+15}_{-19}/-3^f$
& 10.1$^{+1.3}_{-3.2}$ & 0.26$^{+0.02}_{-0.03}$ & 10$^{+33}_{-0p}$ &
1396/1428 \\ 
\end{tabular}
\textit{Notes}: $^f$ parameter fixed at value \\ \phantom{Notes:} $_p$
parameter pegged at lower limit \\ 
\phantom{Notes:} $^\dagger$ second Gaussian added with
$\sigma=0$~\kev. Best fit $E=6.93^{+0.05}_{-0.01}$~\kev, which was
fixed for subsequent fits. \\
\phantom{Notes:} $^\ddagger$ for
these models the cutoff energy=150~\kev\ and abundances=solar \\
\phantom{Notes:} $^\ast$ $r_{\mathrm{out}}$=1000~$r_g$ for the
relativistic models

\end{minipage}
\end{table*}

The fits clearly indicate two distinct emission
lines between 6 and 7~\kev. The lower energy line has an equivalent
width (EW) of 53~eV, is consistent with arising from weakly ionized
iron (Fe~\textsc{i}--Fe~\textsc{xviii}; \citealt{hou69}), and is slightly
broadened (FWHM $\approx 10^{4}$~km~s$^{-1}$). The
unresolved higher energy line at $6.93^{+0.05}_{-0.01}$~\kev\ is
consistent with emission from H-like iron and is not very prominent
(EW$= 13$~eV), but does significantly improve the fit (at $>
99.99$ per cent according to the F-test). These iron line parameters agree
with those presented by \citet{yp04}, who analyzed \textit{Chandra}
High-Energy Gratings (HEG) data of many bright Seyfert~1
galaxies. Those authors also found marginal evidence for an emission
feature at $\sim 6.9$~\kev\ from \threec\, which we confirm with these
\textit{XMM-Newton} data.

Similarly to the previous \textit{BeppoSAX} and \textit{RXTE} analyses
\citep{esm00,zg01,gse03}, the spectrum of \threec\ was well fit with a
moderate amount of reflection.  These high-quality \textit{XMM-Newton}
data constrain the reflection fraction to be $R \sim 0.5$ for a
solar abundance of Fe when using either the PEXRAV \citep{mz95}
reflection model, or the more detailed ionized disc (IONDISC) models
of \citet{rf93} and \citet{rfy99}. Furthermore, the ionization
parameter of the reflector is low, indicating that the prominent
spectral features (the \fe\ line and edge) are dominated by neutral
reflection.

A diskline model for the \fe\ line resulted in a relatively large
value for the disc inner radius ($r_{\mathrm{in}} \approx 200$~$r_g$,
where $r_g=GM/c^2$ is the gravitational radius) for a typical steep
emissivity of $r^{-3}$. However, a similar fit statistic was
found when the inner radius was fixed at $r_{\mathrm{in}}=6$~$r_g$,
and the emissivity allowed to be much flatter. When the curved IONDISC
continuum and line spectrum was blurred with the diskline kernel (a
significant improvement to the non-blurred case; see
Table~\ref{table:fitres}), the value of $r_{\mathrm{in}}$ decreased to
about 75~$r_g$ for an emissivity of $r^{-3}$. The difference is due to
changes in the inclination angle and continuum shape. As mentioned in
Sect.~\ref{sect:intro}, radio data place an upper-limit of 14
degrees to the inclination angle of the jet (and likely the accretion
disc) to the line-of-sight. The minimum angle allowed by the PEXRAV
and PEXRIV codes is 18.2 degrees, and the diskline model was fixed at
that angle for consistency. Allowing the inclination angle to vary in
the PEXRAV model did not improve the fit. As with the PEXRAV+DL model,
an acceptable fit was still obtained when $r_{\mathrm{in}}=6$~$r_g$
and a flatter emissivity allowed.

IONDISC models with iron abundances that differed from solar were also
fit to these data. While the resulting $\chi^2$ was satisfactory, the
model with a supersolar iron abundance was a worse fit than the model
with strictly solar abundances because of problems
accounting for the shape of the iron edge and reflection
hump. However, the IONDISC model with half the solar abundance of iron
yielded the same fit statistic as the solar model by making use of a
steeper primary spectrum of $\Gamma=1.9$ and a higher reflection
fraction.

Extrapolating the PEXRAV+DL+G model to 2~\kev\ yields a
2--10~\kev\ flux of $F_{\mathrm{2-10\ keV}}=4.6\times
10^{-11}$~erg~cm$^{-2}$~s$^{-1}$ and a luminosity of
$L_{\mathrm{2-10\ keV}}=1.2\times 10^{44}$~erg~s$^{-1}$ (assuming a
\textit{WMAP} cosmology: $H_0=70$~km~s$^{-1}$~Mpc$^{-1}$,
$\Lambda=0.73$; \citealt{spe03}). Thus, \threec\ was in a higher flux
state during the \textit{XMM-Newton} observation than found in the
well-studied \textit{ASCA} observation.

\subsection{Broadband models}
\label{sub:broadband}
\begin{figure*}
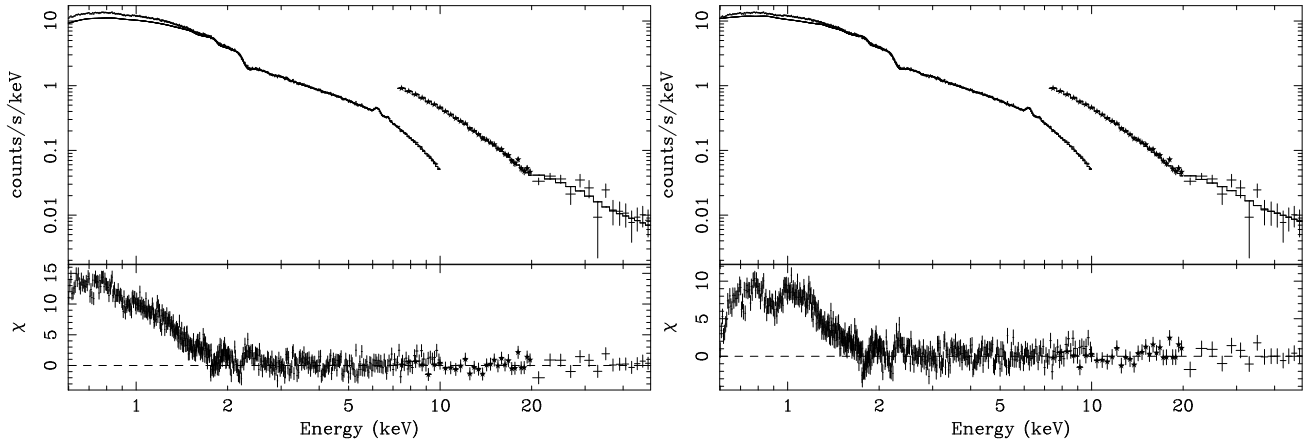

\begin{minipage}{175mm}
\centerline{
\includegraphics[width=0.33\textwidth,angle=-90]{newfig4a.eps}
\includegraphics[width=0.33\textwidth,angle=-90]{newfig4b.eps}
}
\caption{\textit{(Left)}: Count spectra and model residuals (in units
  of standard deviations) when the PEXRAV+DL+G model of
  Table~\ref{table:fitres} is extrapolated to 0.6~\kev. The model
  predicts a significant soft-excess between 0.6 and 2~\kev. \textit{(Right)}:
  Same as the other plot except the blurred 1xFeIONDISC+G model from
  Table~\ref{table:fitres} is extrapolated to 0.6~\kev. A soft excess
  is still predicted, but seems weaker and not as broad as that found
  from the other model. Excess absorption is also predicted. In both figures,
  PCA data are denoted by stars.}
\label{fig:residuals}
\end{minipage}
\end{figure*}

In the previous section, the 3--50~\kev\ spectrum of \threec\ was
found to be well described by a moderate amount of neutral reflection plus two
emission lines. Can this model account for the entire observed X-ray
spectrum of \threec? In Figure~\ref{fig:residuals} we plot the
residuals when two of the best-fitting models from
Table~\ref{table:fitres} (namely, PEXRAV+DL+G and the blurred
1xFeIONDISC+G) are extrapolated to 0.6~\kev.
The plots show that the two models predict very different shapes for
the softer energies. The PEXRAV continuum is very similar to a pure
power-law, and so predicts a significant soft excess (e.g.,
Fig.~\ref{fig:broadband}). The
IONDISC model has a substantial increase in emission at lower energies
due to free-free emission and recombination lines (see, e.g.,
\citealt{rfy99}). As a result, greater than Galactic
absorption is predicted below 0.8~\kev, as well as a
soft-excess. 

\begin{figure}
\centerline{
\includegraphics[width=0.37\textwidth,angle=-90]{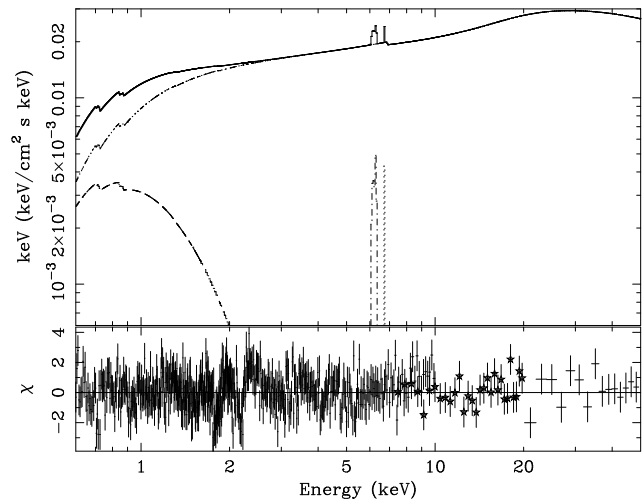}
}
\caption{The upper panel shows the best-fitting broadband model of
  \threec\ using a PEXRAV and bremsstrahlung continuum, plus edges
  from O~\textsc{vii} and O~\textsc{viii}. The solid line
  plots the total model, while the individual curves denote the model
  components. The bottom panel plots the residuals (in
  units of standard deviations). PCA data are denoted by stars.}
\label{fig:pexrav+bremss}
\end{figure}

We first attempted to construct a broadband model of \threec\ by
adding a soft continuum component to the
PEXRAV+DL+G model. Intrinsic absorption was also added to the model in
case it proved necessary. The parameters of the three most successful fits are
shown in Table~\ref{table:softfits}, with the bremsstrahlung continuum
providing the best fit to the soft excess.
\begin{table*}
\begin{minipage}{175mm}
\caption{Parameters resulting from fitting the \threec\ data between
  0.6 and 50~\kev. Intrinsic absorption (using the 'ztbabs' model) and
  various continua models were added to the PEXRAV+DL+G model from
  Table~\ref{table:fitres} to model the residuals shown in
  Fig.~\ref{fig:residuals}. The first column lists the models used to account
  for the soft excess. $N_{\mathrm{H}}$ is the column density of the
  intrinsic absorber (in units of 10$^{21}$~cm$^{-2}$). $kT$
  is the temperature of the fitted blackbody or bremsstrahlung
  component, while $\Gamma_s$ is the photon index of the added
  power-law. $\tau^{\mathrm{max}}_{\mathrm{O\ VII}}$ and
  $\tau^{\mathrm{max}}_{\mathrm{O\ VIII}}$ are the maximum
  optical depths of the O~\textsc{vii} and O~\textsc{viii} edges
  respectively. The other symbols have the same
  meaning as in Table~\ref{table:fitres}. The disc inclination angle
  and diskline emissivity were fixed at 18.2 degrees and $-3$ for all fits. 
 }
\label{table:softfits}
\begin{tabular}{ccccccccccc}
Model & $\Gamma$ & $N_{\mathrm{H}}$ & $E_{\mathrm{b}}$ &
$r_{\mathrm{in}}$ & $R$ & $kT$ & $\Gamma_s$ &
$\tau^{\mathrm{max}}_{\mathrm{O\ VII}}$ &
$\tau^{\mathrm{max}}_{\mathrm{O\ VIII}}$ & $\chi^2/$d.o.f.\\ \hline
bbody & 1.85$\pm 0.01$ & 0$^{+0.01}_{-0p}$ & 6.44$\pm 0.02$ &
175$^{+92}_{-54}$ & 0.59$^{+0.09}_{-0.11}$ & 0.200$\pm 0.004$ & & & & 2040/1911\\
power-law & 1.75$^{+0.02}_{-0.01}$ & 1.4$\pm 0.2$ & 6.44$\pm 0.02$ &
168$^{+74}_{-74}$ & 0.28$\pm 0.10$ & & 4.02$^{+0.21}_{-0.12}$ & & &
2026/1911\\
bremss & 1.82$\pm 0.02$ & 0.28$^{+0.11}_{-0.07}$ & 6.44$\pm
0.02$ & 175$^{+90}_{-53}$ & 0.44$^{+0.12}_{-0.10}$ &
0.46$^{+0.04}_{-0.06}$ & & & & 2024/1911\\
 & 1.83$^{+0.02}_{-0.01}$ & 0.57$^{+0.21}_{-0.10}$ & 6.44$\pm 0.02$ &
182$^{+92}_{-61}$ & 0.45$^{+0.13}_{-0.05}$ & 0.41$^{+0.03}_{-0.04}$ &
& 0.03$^{+0.02}_{-0.01}$ & 0.03$\pm 0.01$ & 1999/1909\\
\end{tabular}
\end{minipage}
\end{table*}
A significant ($>99.99$ per cent) improvement to the broadband
model was found by including absorption edges from O~\textsc{vii} and
O~\textsc{viii}. The maximum optical depths of the edges were measured
with good precision, and are consistent with previous analyses
\citep{rey97,sem99,mck03}. The edges provide strong evidence for warm
absorption along the line-of-sight to \threec\ which will be further
analyzed using RGS data. A plot of this best fitting broadband
model, along with the residuals, is shown in
Figure~\ref{fig:pexrav+bremss}. The remaining residuals shown in the figure are
dominated by systematics in the detector response, such as the neutral
Si K-edge at 1.84~\kev\ and the Au M-edge at 2.2~\kev\
\citep[e.g.,][]{kir03,vf04}. 

The bremsstrahlung model requires only slight intrinsic absorption
with a column of $0.57\times 10^{21}$~cm$^{-2}$. This result was from
a cold absorption model, but with the strong evidence for warm
absorption provided by the oxygen edges, the actual value of the
intrinsic cold absorbing column will likely be lower. Absorption in excess
of the Galactic column was also inferred from \textit{ROSAT}
\citep{gr97}, \textit{ASCA} \citep{rey97,sem99}, and \textit{BeppoSAX}
\citep{zg01} data. However, it was not found in \textit{EXOSAT}
data \citep{mar91} or in the recent \textit{Chandra} observation
\citep{mck03}, although this would depend on the correction for the
ACIS degradation. 

As seen in Figure~\ref{fig:pexrav+bremss}, the
bremsstrahlung component does not dominate the overall X-ray flux from
\threec, contributing only 3.5 per cent of the 0.6--50~\kev\ flux
($1.38\times 10^{-10}$~erg~cm$^{-2}$~s$^{-1}$) and 20 per cent of the
0.6--2~\kev\ flux ($2.31\times 10^{-11}$~erg~cm$^{-2}$~s$^{-1}$). The
EWs of the emission lines are 53~eV for the broad \fe\ line and 11~eV
for the ionized line, very similar to the values obtained from the
3--50~\kev\ fits.

Intrinsic absorption, oxygen edges and soft excess models were then
added to the IONDISC continuum model (Fig.~\ref{fig:residuals},
right-hand side), but the fits lost the good results at the high
energy end.  This is because the neutral
reflection models predict Ly$\alpha$ emission lines from H-like Mg
and Si between 1 and 2~\kev, as well as other recombination lines at
lower energies \citep[e.g.,][]{rfy99}.  These emission lines are not
required by the data at their predicted strengths (for solar
abundances), and the fit therefore dropped
the reflection fraction by $\sim 2$, resulting in a poorer description
to the \fe\ line and reflection hump. One way to diminish the
importance of the softer lines in the IONDISC model is if the
reflection spectrum is a result of reprocessing from material at some
distance from the illuminating power-law. Then the reflection spectrum
can be adjusted independently of the continuum. Indeed, the values of
$R$ and $r_{\mathrm{in}}$ found by fitting the 3--50~\kev\ spectrum
indicate that the reprocessing is occurring well away from the central
engine (Table~\ref{table:fitres}).

We therefore attempted to fit the 0.6--50~\kev\ spectrum of \threec\
with a model of a power-law continuum and distant reflection (modeled
with the IONDISC code), as well as intrinsic absorption,
O~\textsc{vii} and O~\textsc{viii} edges, and a bremsstrahlung
spectrum to account for the soft excess. This model is not too
dissimilar from the two component model of MCG--6-30-15 put forward by
\citet{fv03} and \citet{mf04}. In this case, however, the reflection
component does not extend to the innermost stable orbit of the
accretion disc. The results of the spectral fit with this model are
shown in Table~\ref{table:softfits2}. 
\begin{table*}
\begin{minipage}{172mm}
\caption{Parameters resulting from fitting the \threec\ data between
  0.6 and 50~\kev. Intrinsic absorption (using the 'ztbabs' model), a
  bremsstrahlung spectrum, O~\textsc{vii} and O~\textsc{viii} edges,
  and an additional continuum component were added to the blurred
  1xFeIONDISC+G model, where the IONDISC spectrum is now reflection
  dominated. The first column lists the added broadband continuum
  model. $N_{\mathrm{H}}$ is the column density of the intrinsic
  absorber (in units of 10$^{21}$~cm$^{-2}$). $kT$ is the temperature
  of the bremsstrahlung component that fits the soft excess, and
  $\xi_{\mathrm{r}}$ is the ionization parameter of the
  reflection-dominated reprocessor. The other symbols have the same
  meaning as in Tables~\ref{table:fitres} and
  \ref{table:softfits}. The emissivity and outer radius were fixed at
  $-3$ and $1000$~$r_g$, respectively, for all fits.
 }
\label{table:softfits2}
\begin{tabular}{ccccccccccc}
Added Continuum & $\Gamma$ & $N_{\mathrm{H}}$ & $\xi_{\mathrm{r}}$ &
$r_{\mathrm{in}}$ & $i$ & $\log \xi$ & $kT$ &
$\tau^{\mathrm{max}}_{\mathrm{O\ VII}}$ & $\tau^{\mathrm{max}}_{\mathrm{O\ VIII}}$ & $\chi^2/$d.o.f.\\ \hline
Power-law & 1.86$\pm 0.01$ & 1.0$^{+0.2}_{-0.1}$ & 10$^{+0.2}_{-0p}$ &
66$^{+25}_{-15}$ & 9.7$^{+1.4}_{-4.5}$ & & 0.37$\pm 0.02$ & 0.06$\pm
0.01$ & 0.04$\pm 0.01$ & 2034/1909\\
Ionized Reflector$^\dagger$ & 1.74$\pm 0.02$ & 1.2$^{+0.1}_{-0.2}$ &
10$^{+0.3}_{-0p}$ & 70$^{+24}_{-17}$ & 9.3$^{+2.8}_{-5.0}$ &
5$^{+0p}_{-0.14}$ & 0.31$\pm 0.01$ & 0.07$\pm 0.01$ & 0.03$\pm 0.01$ &
2027/1908\\
\end{tabular}
\textit{Notes}: $^\dagger$ highly ionized reflector added in with $R$
fixed at unity\\
\phantom{Notes:} $^p$$_p$ parameter pegged at upper/lower limit \\  
\end{minipage}
\end{table*}
A decent fit is obtained, but with a worse reduced $\chi^2$ as was found
for the PEXRAV+DL and bremsstrahlung model
(Table~\ref{table:softfits}). Again, this is due to the presence of
the soft emission lines predicted by the model. However, these lines must be
there at some level in a neutral reflection spectrum, so the better
fit by the PEXRAV continuum argues for a subsolar abundance of metals
(e.g., O, Mg, Si, and Fe) in \threec. Recall that the 3--50~\kev\ spectrum
was equally as well fit by a IONDISC model with half the solar
abundance of iron. Unfortunately, grids of IONDISC models with other
metal abundances differing from solar are not available, so their
values cannot be directly constrained. Yet, there is optical spectroscopic
evidence for non-solar abundances within \threec. Both \citet{bald80}
and \citet{fm80} find that the metal abundance (in particular O) in
the nebular gas may differ from solar by factors of several. It is
reasonable to expect that the gas feeding onto the black hole will not
have a too dissimilar abundance pattern.

\begin{figure}
\centerline{
\includegraphics[width=0.37\textwidth,angle=-90]{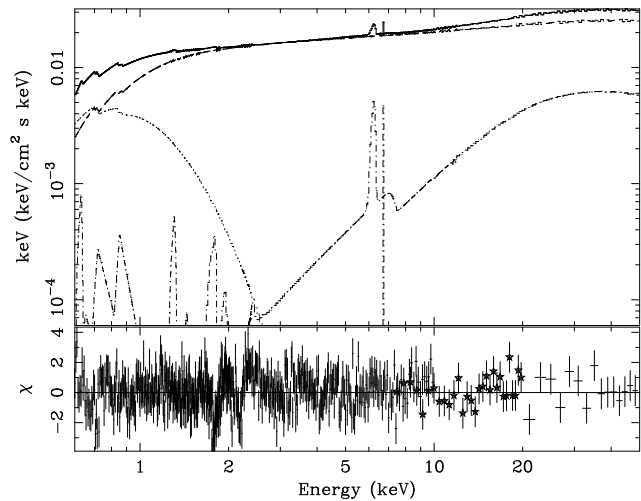}
}
\caption{The upper panel shows a broadband model of
  \threec\ using two 1xFeIONDISC models and a bremsstrahlung
  continuum. The solid line plots the total model, while the
  individual curves denote the model components. The bottom panel
  plots the residuals (in units of standard deviations). PCA data
  are denoted by stars.}
\label{fig:reflectmodel}
\end{figure}

An improvement to the fit was obtained when the additional
power-law was replaced by a highly-ionized reflector (from the same
set of IONDISC models). The added reflection component included the
intrinsic power-law, but had $R$ fixed at unity. This model is similar
to the two-component ionized disc fit to MCG--6-30-15 presented by
\citet*{bvf03}, but this new reflection component is ionized to the
point that it does not contribute much, if any, emission to the \fe\
line or absorption to the iron edge (Figure~\ref{fig:reflectmodel}).
It does however exhibit a small amount of spectral curvature due to
free-free emission at low energies from the hot surface of the
slab. This small amount of curvature in the spectrum obviously makes
an improvement to the fit from the power-law spectrum
($\Delta \chi^2=7$ for 1 additional degree of freedom, giving a $F$-test
probability of 98.9 per cent). Since this
ionized spectrum does not predict a significant spectral feature, it
could not be relativistically blurred to constrain an emission
radius. It seems reasonable to expect that it arises from the
innermost parts of the accretion disc \citep[e.g.,][]{bvf03}. The EWs
of the two emission lines are 50~eV and 12~eV, in good agreement with
those obtained from the 3--50~\kev\ fits (Sect.~\ref{sub:hardxrays}).

As expected from Fig.~\ref{fig:residuals}, the intrinsic absorption
column is larger for the IONDISC-based models than with the
PEXRAV-based ones, with a value of $N_{\mathrm{H}} \approx 1\times
10^{21}$~cm$^{-2}$. Again, this values will likely drop by a small
amount once the warm absorber has been correctly modeled.

The best fitting bremsstrahlung temperature is an order of magnitude
lower than the surface temperature of the ionized reflector, and
contributes only 0.2 per cent of the total 2--10~\kev\ flux, $F_{\mathrm{2-10\
    keV}}=4.65\times 10^{-11}$~erg~cm$^{-2}$~s$^{-1}$. However, it
produces 23 per cent of the total 0.6--2~\kev\ flux, $F_{\mathrm{0.6-2\
    keV}}=2.31\times 10^{-11}$~erg~cm$^{-2}$~s$^{-1}$. 

The reflection dominated spectrum has a very low ionization parameter,
so is dominated by a neutral \fe\ line plus recombination lines at
lower energies. The width of the \fe\ line restricts the reflection
region to be greater than $\approx 70$~$r_g$ from the black hole (of
course, a flatter emissivity would allow a lower $r_{\mathrm{in}}$). This
component contributes 3 per cent of the total 2--10~\kev\ flux and 9
per cent of the total 0.6--50~\kev\ flux, $1.41\times
10^{-10}$~erg~cm$^{-2}$~s$^{-1}$.

Using the PEXRAV+DL+G and bremsstrahlung continuum from
Table~\ref{table:softfits}, we find that the total unabsorbed X-ray
luminosity between 2 and 50~\kev\ is $L_{\mathrm{2-50\
keV}}$=2.9$\times 10^{44}$~erg~s$^{-1}$ ($L_{\mathrm{2-10\
keV}}$=1.2$\times 10^{44}$~erg~s$^{-1}$;
Sect.~\ref{sub:hardxrays}). The Eddington luminosity for a $3\times
10^7$~M$_{\odot}$ black hole is $3.78\times 10^{45}$~erg~s$^{-1}$, so
if the hard X-rays contribute about 10 per cent of the bolometric
luminosity of \threec\ \citep{elv94}, then this observation caught the
source accreting very close ($\sim 77$ per cent) to the Eddington
rate. However, given the small inclination angle into the central
engine, it is possible that the observed flux is enhanced due to
beaming along the line-of-sight.

\subsection{An earlier \textit{XMM-Newton} observation of \threec}
\label{sub:previous}
\threec\ was previously observed by \textit{XMM-Newton} during
revolution 502 as part of a Guaranteed Time observing program. The
observation occurred on 2002 September 6, and lasted for 12.6 ks. The
EPIC MOS2 and pn were in small-window mode, and the MOS1 detector was
in timing mode. The ODF was downloaded from the \textit{XMM-Newton}
Science
Archive\footnote{http://xmm.vilspa.esa.es/external/xmm\_data\_acc/xsa/index.shtml},
and reduced using SAS v.6.0. following the same procedures as
described in Sect.~\ref{sub:xmm}. Below, we analyze the pn spectrum
from this observation. The total good exposure time after filtering
and extraction was 8250 s. These data caught \threec\ in a slightly
lower flux state ($F_{\mathrm{2-10\ keV}}=4.0\times
10^{-11}$~erg~cm$^{-2}$~s$^{-1}$), and so it interesting to try our
best fitting time-averaged models from the previous section to see if
there were any important changes to the fit parameters.

We again fit the pn data from 0.6--10~\kev, and used two models: (1) the
bremsstrahlung and PEXRAV model from Table~\ref{table:softfits}, and (2)
the bremsstrahlung+2 IONDISC model from
Table~\ref{table:softfits2}. Intrinsic absorption and O~\textsc{vii}
and O~\textsc{viii} edges were also included in the models. The
results are shown in Table~\ref{table:oldfits}.
\begin{table*}
\begin{minipage}{175mm}
\caption{Parameters resulting from fitting the archival \threec\ data
  between 0.6 and 10~\kev. The model included intrinsic absorption
  (using the 'ztbabs' model), O~\textsc{vii}
and O~\textsc{viii} edges, a bremsstrahlung spectrum, plus one of
  the two different best fitting continuum models from the previous
  section. Model \#1 is the
bremsstrahlung and PEXRAV model from Table~\ref{table:softfits}, while
  model \#2 is the bremsstrahlung+2 IONDISC model from
Table~\ref{table:softfits2}. All symbols have the same meaning as in previous
  tables. EW$_1$ is the equivalent width (in eV) of the neutral \fe\
  line. EW$_2$ is the equivalent width (in eV) of the narrow, ionized
  line. The diskline emissivity and outer radius was fixed at $-3$ and
  $1000$~$r_g$, respectively, for all fits.
 }
\label{table:oldfits}
\begin{tabular}{cccccccccccc}
Model & $\Gamma$ & $N_{\mathrm{H}}$ & $E_{\mathrm{b}}/\xi_{\mathrm{r}}$ &
$\sigma_{\mathrm{b}}/r_{\mathrm{in}}$ & $R/\log \xi$ & EW$_1$ &
$kT$ & EW$_2$ & $\tau^{\mathrm{max}}_{\mathrm{O\ VII}}$ &
$\tau^{\mathrm{max}}_{\mathrm{O\ VIII}}$ & $\chi^2/$d.o.f.\\ \hline
1$^\dagger$ & 1.90$^{+0.08}_{-0.14}$ & 0.49$^{+0.82}_{-0.48}$ &
6.43$^{+0.10}_{-0.09}$ & 0.09$^{+0.15}_{-0.09p}$ & 1.9$^{+1.1}_{-0.5}$
& 40 & 0.33$^{+0.11}_{-0.06}$ & 2 & 0.01$^{+0.06}_{-0.01p}$ & 0$^{+0.04}_{-0p}$ & 1235/1232\\
2$^\ast$ & 1.60$^{+0.09}_{-0.08}$ & 0.82$^{+0.63}_{-0.67}$ &
10$^{+4}_{-0p}$ & 80$^{+919p}_{-48}$ & 4.98$^{+0.02p}_{-1.02}$ & 55
& 0.34$^{+0.14}_{-0.06}$ & 16 & 0.04$^{+0.05}_{-0.04p}$ &
0$^{+0.04}_{-0p}$ & 1245/1232\\
\end{tabular}
\textit{Notes}: $^\dagger$ G replaced DL since $r_{\mathrm{in}}$ unconstrained in DL model\\
\phantom{Notes:} $^\ast$ highly ionized reflector has fixed $R=1.0$;
other reflector has inclination angle frozen at 9.3 degrees\\ 
\phantom{Notes:} $^p$$_p$ parameter pegged at upper/lower limit \\  
\end{minipage}
\end{table*}
Both models successfully fit the data with similar values of
the reduced $\chi^2$, but point to two different interpretations
of the spectrum. In both cases, the oxygen absorption edges are not
required by the fits, but the upper-limits are consistent with the
values found from the later, longer \textit{XMM-Newton} dataset
(similarly for the intrinsic absorption columns).

The PEXRAV model results in a steep $\Gamma \approx 1.9$ spectrum
which has a high reflection fraction of $R \sim 2$. The neutral \fe\
line is still present, but now has a reduced EW of 40~eV. These
measurements imply a drastic change in the source between the two
observations, even though the flux differed by only $\sim 13$ per
cent. Indeed, a steeper spectrum at this flux state would be unusual,
as \threec\ has been observed to follow the typical Seyfert behaviour
of exhibiting a harder spectrum at lower count rates
\citep[e.g.,][]{gse03}. The very strong Fe edge in the model
diminished the contribution of the weak ionized line at $\sim
6.9$~\kev.

In contrast, the IONDISC model results in a lower $\Gamma$ at this
flux state, in agreement with past observations. The reduced $\chi^2$
is slightly worse than with the PEXRAV model, but, as in
Sect.~\ref{sub:broadband}, this can be attributed to overestimating
the metal abundances in the source. A stronger \fe\ edge is possibly
needed for this dataset, as ionization parameters that are smaller by
a factor of a few are allowed for the highly ionized reflector. Lower
values of $\xi$ will produce a more prominent (but still broad;
\citealt{rfy99}) iron edge. Unfortunately, the inner radius of the
distant reflector is unconstrained at large radii, but is consistent
with the neutral reflector being brought closer to the central engine.
The strength of this component in the 2--10~\kev\ band, relative to
the total, was the same 3 per cent as in the longer observation, but
the total flux in the bremsstrahlung component was lower by 18 per
cent. The total unabsorbed luminosity between 2 and 10~\kev\ predicted
by this model is $L_{\mathrm{2-10\ keV}}$=$10^{44}$~erg~s$^{-1}$.

\section{Timing analysis}
\label{sect:timing}
\begin{figure}
\centerline{
\includegraphics[width=0.37\textwidth,angle=-90]{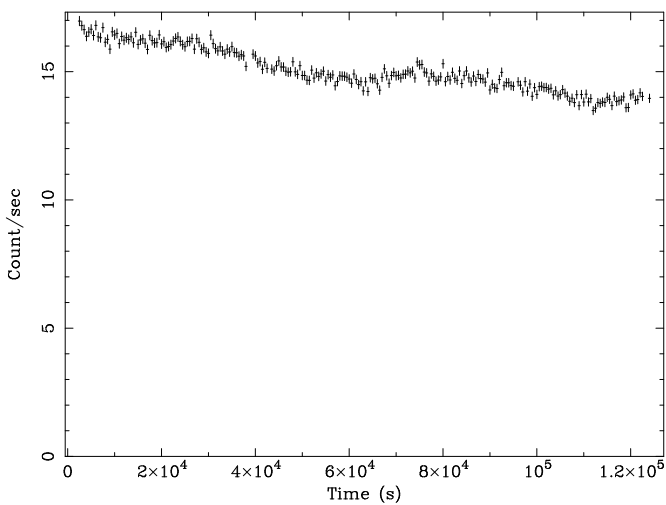}
}
\caption{EPIC pn 0.2--10~\kev\ lightcurve of \threec\ in 500~s
  bins. The source shows small short
  timescale variability, and a decreasing trend from the beginning to
  the end of the observation.}
\label{fig:lightcurve}
\end{figure}

\threec\ is a known variable X-ray source \citep[e.g.,][]{m04},
although previous observations have shown that any large amplitude
variations occur on timescales of days to weeks \citep{gse03} rather
than on the hours-to-days timescale of pointed observations
\citep{rey97,mck03}. Such is the case with the current
\textit{XMM-Newton} observation. Figure~\ref{fig:lightcurve} plots the
0.2--10~\kev\ EPIC pn lightcurve of \threec\ in 500~s bins.
The lightcurve is dominated by a long timescale declining trend as the
source count-rate drops by $\sim 20$ per cent from the start to the end of
the observation. Fitting the lightcurve with a constant yields
$\chi^2/$d.o.f.=5493/238. As it decays \threec\ also exhibits low
amplitude flickering on the order of 2--3 per cent.

\begin{figure}
\centerline{
\includegraphics[width=0.37\textwidth,angle=-90]{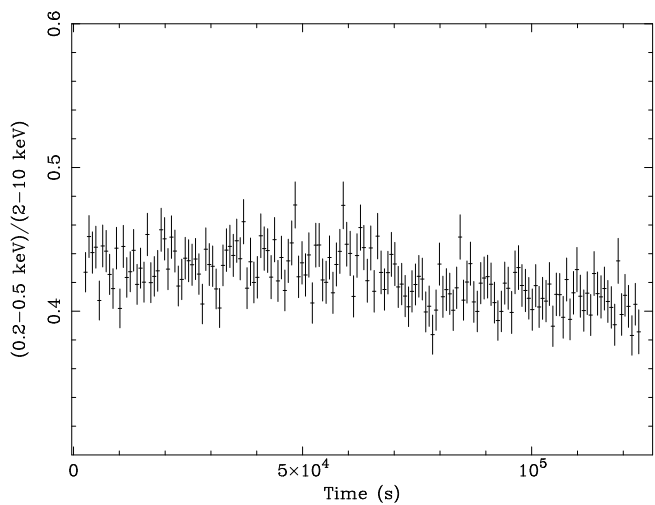}
}
\caption{The EPIC pn (0.2--0.5~\kev)/(2--10~\kev) softness ratio of \threec\
  as a function of time (in 750~s bins) during the \textit{XMM-Newton}
  observation. The source shows both long
  and short timescale mild spectral variability.}
\label{fig:sr}
\end{figure}

Evidence for spectral variability has been previously presented by
\citet{zg01} and \citet{gse03}. These authors found that \threec\ was
more variable at lower energies, and showed the typical Seyfert~1
characteristic of becoming softer at higher
count-rates. Figure~\ref{fig:sr} plots the EPIC pn 0.2--0.5~\kev\ to
2--10~\kev\ softness ratio from the long \textit{XMM-Newton} observation.
Modest spectral variability is observed in the source (a constant fit
results in $\chi^2/$d.o.f.=240/162), but it is not in the
straightforward manner described above. It appears that the softness
ratio remains roughly constant over the first 60~ks of the observation
even though the overall count-rate is falling
(Fig.~\ref{fig:lightcurve}). Beyond this point, the source shows a
gradual hardening. The amplitude of the decrease in the softness ratio
is small, which may be the result of the relative minor change in
count-rate. Observations that probe longer timescales
\citep[e.g.,][]{gse03} are necessary to fully understand the
variability properties of \threec.

\begin{figure}
\centerline{
\includegraphics[width=0.45\textwidth]{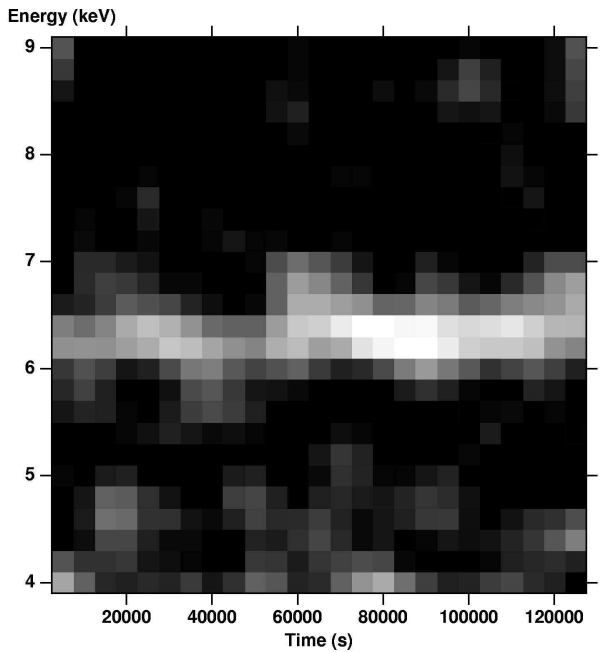}
}
\caption{The image plots the excess flux above the best-fit power-law
  continuum as a function of both energy (in 0.2~\kev\ bins) and time
  (5~ks resolution). The colour map ranges linearly from
  0~ph~cm$^{-2}$~s$^{-1}$ (black) to $1.92\times
  10^{-5}$~ph~cm$^{-2}$~s$^{-1}$ (white). The plot
  shows an apparent increase (significant at the $\sim 90$ per cent level)
  in the excess flux in the observed \fe\ band at $\sim 80$~ks into
  the observation. Full details on the construction of such images
  will be described by Iwasawa \etal\ (in prep).}
\label{fig:image}
\end{figure}

\begin{figure}
\centerline{
\includegraphics[width=0.37\textwidth,angle=-90]{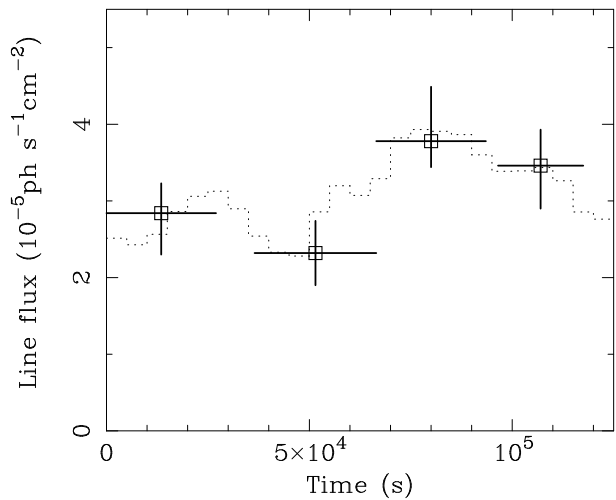}
}
\caption{The points show the \fe\ line flux from four different time
  intervals in the EPIC pn lightcurve. The error-bars show the 1$\sigma$
  uncertainty in the measurement. The 90 per cent error
bars of the second and third data points do not overlap. The dotted
  line shows the smoothed 5 ks resolution Fe K lightcurve from the
excess image (Fig.~\ref{fig:image}).}
\label{fig:linefluxlc}
\end{figure}

There is also modest evidence for an increase in the \fe\ line
flux in the latter part of the observation. Figure~\ref{fig:image} shows
an image plotting the excess flux above the best-fit power-law
continuum as a function of both energy (between 4 and 9~\kev) and
time. For full details on how the image was constructed see Iwasawa
\etal\ (in prep). Concentrating only on the observed \fe\ line band,
we see that the excess flux appears to increase at $\ga 80$~ks into
the observation, before diminishing 10--20~ks later. The observed increase is
significant at $\sim 90$~per cent level (Figure~\ref{fig:linefluxlc}),
so future observations are required to verify the possibility of a
variable \fe\ line in \threec.

\section{Discussion}
\label{sect:discuss}
In the previous sections we have presented the results of a
simultaneous 127~ks \textit{XMM-Newton} and \textit{RXTE} observation
of \threec, the brightest BLRG at X-ray energies. This dataset
provides the highest quality 0.6--10~\kev\ spectrum of a radio-loud
AGN ever produced. Below, we will discuss the results of our analysis
and attempt to place then in the larger context of AGN and jet
physics.

\subsection{Accretion mode}
\label{sub:accretion}
One of the long-standing puzzles in the study of AGN is the cause of
the wide range of radio powers produced by the central engine. Only
about 15--20 per cent of quasars and Seyfert galaxies produce large-scale,
luminous radio jets \citep{kel89}. These radio-loud AGN tend to dwell
in large elliptical galaxies, and thus have central black holes with
masses $\geq 10^8$~M$_{\odot}$ \citep{lao00,md01,jm02}. In contrast,
the radio-quiet AGN mostly inhabit spiral galaxies and so have less
massive black holes. However, recent compilations of AGN black hole
masses suggest that radio-loud sources can exist over a wide range of
black hole mass and accretion rate (\citealt{ho02,oww02,wu02}; for a
dissenting view see \citealt{lao04}). Indeed, low-luminosity
AGNs, such as LINERs, also appear to be more frequently radio-loud
than Seyfert galaxies \citep{ho02,nfwu02}.

As mentioned in Sect.~\ref{sect:intro}, previous observations of
\threec\ and other BLRGs by \textit{ASCA} or \textit{RXTE} have
found slightly harder X-ray continua and weaker reflection
features than the radio-quiet Seyfert~1s \citep[e.g.,][]{eh98}. These
facts favoured an interpretation of an accretion geometry where the
optically-thick disc was truncated into an optically-thin, but
geometrically thick flow. Our analysis of the \textit{XMM-Newton} data
of \threec\ confirms the previous findings of apparent weak reflection
and \fe\ line (Table~\ref{table:fitres}). Moreover, the neutral \fe\
line is observed to be rather narrow (FWHM$\sim 10^4$~km~s$^{-1}$),
although, because of the low inclination angle into the system, this
corresponds to a disc radius of $\sim 75$~$r_g$ assuming a $r^{-3}$
emissivity. Based on these facts alone it would be tempting to
conclude that this is strong evidence for a truncated accretion flow
in \threec. However, if a flatter emissivity profile is allowed, then
the line may extend down to the innermost stable circular orbit (ISCO)
for a Schwarzschild black hole with little change in the fit statistic
(Table~\ref{table:fitres}). Such flat emissivity profiles have been
postulated to explain some of the more unusual narrow lines observed
by \textit{XMM-Newton} \citep{yaq03,vau04}. While there is no
\textit{a priori} reason to expect flat emissivity profiles in the
central engines of AGN, it is important to realize that the geometry
of the X-ray emitting region is unknown even in radio-quiet
Seyfert~1s. Since radio-loud AGN are already unusual in the manner of
their radio emission, one could not rule out a non-standard emissivity
profile for the \fe\ line. For example, power removed from the disc by
the jet could reduce the emissivity profile near the center. If this
power is then released as an X-ray source $\sim 100$~$r_g$ above the
black hole \citep[e.g.,][]{ghm04}, a flat emissivity profile will also
result \citep{vau04}. Alternatively, the jet may obscure the central
regions of the disc from view, resulting in a narrower iron line (this
could be relevant for face-on sources such as \threec). Thus, the width
of the iron line in the \textit{XMM-Newton} data cannot be used to
argue for a truncated accretion disc in \threec. The low reflection
amplitude can be explained by a subsolar iron abundance
(Table~\ref{table:fitres}) or dilution from another continuum
component, rather than a low intrinsic covering factor. As all three
explanations result in roughly the same statistical fit to the data,
there is no clear indication for a change in accretion geometry in
\threec.

An alternative explanation for the low reflection fraction and weak
\fe\ line in BLRGs was put forward by \citet{brf02}. These authors
suggested that the observed properties could be explained by reflection
from a highly photoionized accretion disc that may extend unaltered
down to the ISCO. This model can fit the \textit{XMM-Newton} data
presented here, but requires a double reflection model
\citep[cf.,][]{bvf03} where the innermost reflector is highly ionized,
and a neutral outer reflector which produces the observed
\fe\ line (see Fig.~\ref{fig:reflectmodel}). Moreover, subsolar metal
abundances (e.g., O, Mg and Si) are also required in order for the
emission features at softer energies to be consistent with the
data. However, the inner ionized reflector was required to have a very
high ionization parameter, so that the reflection spectrum, when
combined with the incident power-law, was almost indistinguishable
from a power-law itself. As a result, we were unable to use
relativistic blurring to test if it was consistent with arising from
close to the ISCO. Furthermore, the differences between the highly
ionized reflector and a pure power-law which resulted in the better
fit are subtle and thus are dominated by systematics in the data
and response matrices rather than physics. Therefore, although the
double reflection model provided almost as good a spectral fit, it
also cannot be judged to be the preferred accretion geometry for
\threec.

The 12.7~ks observation of \threec\ caught the source in a slightly
lower flux state than the long \textit{XMM-Newton} pointing, so one
would hope to see evidence of a lower ionization state in the inner
accretion disc. When fit to these data, the double reflection model
provided a good fit with an ionization parameter and inner radius
consistent with being smaller, but with large error-bars. It seems
that to truly test this model (as well as the others) would require
fits at a number of different flux levels. As \threec\ varies most
strongly on timescales of days to weeks \citep{gse03,m04}, a number of
20--30~ks observations each separated by a month or two, might be the
best strategy for finally determining the accretion mode in \threec.

\subsection{The role of the relativistic jet}
\label{sub:jet}
\threec\ possesses a superluminal radio jet which constrains the
inclination angle to $< 14$ degrees. The angle resulting from the
relativistic fits to the X-ray spectrum is consistent with this limit
(Tables~\ref{table:fitres} and \ref{table:softfits2}). This low
inclination angle implies that \threec\ may show some blazar
characteristics, such as rapid X-ray variability or a very hard
spectrum. At the very least, a jetted X-ray component could dilute a
Seyfert spectrum, and cause the observed weak reflection features. The
\textit{XMM-Newton} observation clearly shows that any blazar-like
features due to the jet do not dominate the X-ray emission. The \fe\
line, reflection hump, spectral-energy distribution \citep{mar91}, and
lack of COMPTEL or EGRET detection \citep{mai95,vm95} all point to a
high-energy spectrum that is Seyfert-like
\citep[e.g.][]{zg01}. Furthermore, the observed variability properties
\citep{gse03,m04} are also inconsistent with a jet origin for the
X-ray emission. While it is hard to place a strict limit on any
contribution by a blazar-like component (\citet{gse03} estimate 5 per cent),
the \textit{XMM-Newton} data do not provide any evidence for this
component. A higher-energy observation by \textit{Integral} or
\textit{ASTRO-E2} is necessary to determine the amount, if any, of jet
contamination to the X-ray emission of \threec.

\subsection{The origin of the soft excess}
\label{sub:soft}
The \textit{XMM-Newton} data indicates a clear need for a soft excess in the
spectrum of \threec, in agreement with earlier work
\citep{wc92,gr97,zg01}. However, we found that the soft excess is best
fit by a thermal bremsstrahlung spectrum rather than a power-law
\citep{gr97} or a line-emitting plasma \citep{zg01}. The best fit
bremsstrahlung model had a temperature of $kT \approx 0.3$--$0.4$~\kev\ or
$3.5$--$4.6\times 10^6$~K, and was required to be subject to intrinsic as
well as Galactic absorption (as opposed to the plasma model of
\citealt{zg01}). This agrees with the fact that the zeroth-order
\textit{Chandra} image of \threec\ did not detect a significant
extended component \citep{mck03}. The total unabsorbed luminosity in
the bremsstrahlung spectrum is $1.6\times 10^{44}$~erg~s$^{-1}$, which is
less than half of the total X-ray luminosity of \threec,
$L_{\mathrm{0.6-50\ keV}}=3.8\times 10^{44}$~erg~s$^{-1}$. The unabsorbed
free-free spectrum peaks in $EF_E$ space at $\sim$0.2~\kev.

The hot temperature of the bremsstrahlung spectrum and the close
relation between the luminosities strongly suggests that the X-ray
emission is the cause of the hot emitting gas. The lower
bremsstrahlung flux observed from the archival observation of \threec\
also points to a connection between the free-free emitting gas and the
AGN. Two possible origins for this hot gas are the broad-line region,
and nearby giant H~\textsc{ii} regions. Optical studies of the broad
emission lines in \threec\ have all noticed the presence of strong He
lines \citep[e.g.,][]{mol88}, which can be explained if the lines
form in regions of very high density ($n_e \ga 10^{10}$~cm$^{-3}$;
\citealt{rud87}). The inclination angle into \threec\ is such that our
line-of-sight may pass through a significant volume of the
photoionized gas that makes up the broad-line region. With the high
density implied by the He emission lines, a strong free-free continuum
may be emitted.  This scenario could be tested by other observations
of Seyferts or BLRG with low inclination angles.

The area directly around the nucleus of \threec\ is a chaotic region
of photoionized gas \citep{hua88,sou89}, with plenty of H$\alpha$ and
O~\textsc{[iii]} emission observed to be within the point-spread function of
\textit{XMM-Newton} \citep{hua88}. If the AGN dominates the ionization
of this surrounding gas, then, assuming the presence of high density gas to
increase the free-free emissivity, these giant H~\textsc{ii}
regions may be the source of the observed bremsstrahlung component. A
detailed analysis of the \textit{Chandra} data which would miss most
of this emission could test this model.

\subsection{Implications for understanding jet formation}
\label{sub:jetformation}
The goal of X-ray spectroscopy of radio-loud objects is to elucidate
details on how relativistic collimated outflows are generated by
comparing the results to the more numerous radio-quiet sources. Has
this \textit{XMM-Newton} observation of \threec\ brought us closer to
that goal? The exact accretion geometry close to the central black
hole remains uncertain (Sect.~\ref{sub:accretion}). The observed luminosity
and inferred black hole mass imply that \threec\ is accreting close to
its Eddington rate. This would seem to make a truncated accretion disc
geometry rather unlikely if the inner disc had to take on a
radiatively-inefficient form such as an ADAF (advection dominated
accretion flow; \citealt{ny95}), which is stable only at low accretion
rates. At high accretion rates, the disc will likely be thicker due to
radiation pressure (see, e.g., 'slim' discs; \citealt{aba88}), but can
extend down to the ISCO. If radio-loudness is more common
\citep{xlb99} at very high and very low luminosities (where ADAFs are
more likely) then that would suggest that thick accretion flows may be very
important for the launching of jets (see also \citealt{mcf04} and
\citealt{bc04}). This is consistent with some of
the more recent theoretical considerations \citep{mei99,mei01} which
notes the importance of the poloidal magnetic field in the disc for jet
production.

A highly ionized accretion disc would be consistent with a high accretion
rate. Ionized \fe\ lines have been observed in \textit{XMM-Newton}
observations of the high-luminosity Seyferts Mrk~205 and Mrk~509
(\citealt{ree01,pou01,der04}; but see \citealt{pds03}). Ionized discs have long been considered an
explanation for the weak-to-non-existent \fe\ lines in quasars (i.e.,
the X-ray Baldwin effect; \citealt{iw93,nan97b}), and have been
successfully applied to some narrow-line Seyfert~1s (NLS1s) which are
believed to be accreting at a significant fraction of Eddington
\citep{bif01,long03,pou03}. Interestingly, there are very few known
radio-loud NLS1s which suggests that a high or low accretion rate
relative to Eddington is not the only parameter important for jet
formation. NLS1s have lower than typical black hole masses
\citep{fer01,whd04}, which means their accretion discs would likely be
denser and not as thick as ones around higher mass black holes
accreting at the same fraction of Eddington \citep{ss73}. If jet
production is strongly sensitive to disc thickness then this would be
a possible explanation for apparent correlation between radio-loud AGN
and black holes with masses $\ga 10^7$~M$_{\odot}$. These ideas can
also be extended down to the lower mass Galactic black hole
candidates, where jet production is often observed only in the
low/hard state \citep{gfp03}, and is quenched when the systems move to
the high/soft state \citep[e.g.,][]{fen99}. A truncated accretion disc
geometry in the low/hard state when the accretion rate is small,
moving to a thinner and denser configuration in the more rapidly
accreting high/soft state would be consistent with the picture
described above. Clearly, more observations are required to test the
above scenario. In this regard, ionized disc modeling will be vitally
important; as we have seen with \threec, there can be very subtle
differences between a power-law and ionized-disc continuum which will
need to be taken into account in order to obtain an accurate picture
of the inner accretion disc in AGN.

\section{Summary and Conclusions}
\label{sect:concl}
This paper presented EPIC data from a 127~ks \textit{XMM-Newton}
observation of \threec\ that was performed simultaneously with
\textit{RXTE}. The time averaged spectral fitting found similar
results to previous observations: a weak reflection spectrum, with a
relatively small (EW$\approx 50$~eV) neutral \fe\ line. The data
strongly requires absorption greater than the Galactic value and a soft
excess. Small absorption edges from O~\textsc{vii} and O~\textsc{viii} also
significantly improve the fits, implying the presence of a weak warm
absorber. The soft excess was best fit by a thermal bremsstrahlung component,
which we speculate arises from either the broad-line region or
photoionized regions close to the nucleus. The data also prefers a weak
ionized iron line at $\sim 6.9$~\kev, as first suggested by a
\textit{Chandra} observation \citep{yp04}. Tentative evidence for \fe\
variability was presented which needs to be verified by future
observations.

We are unable to convincingly determine whether a truncated or
non-truncated accretion disc geometry provides the best description of
the data. A basic power-law plus simple distant reflector model
provided the best statistical fit.  A double reflection model,
combining the features of a very highly ionized and a neutral
reflector also fits the broadband data, but requires subsolar
metal abundances based on the strength of the soft X-ray emission
features. There are no constraints on the location of the highly
ionized reflector, but it most plausibly arises from very close to the
ISCO. The neutral reflector arises from the outer parts ($\ga
75$~$r_g$) of the disc. The spectral and variability results confirm
the source is Seyfert-like, and that any jet emission in the observed
band will be very small.

This observation has clarified many of the outstanding questions
regarding the spectrum of \threec, but unfortunately further
observations are required to produce a convincing physical model. The
fact that \threec\ was accreting close to the Eddington limit is
consistent with the ionized disc model and implies that disc thickness
may be a key parameter in the ability of a system to launch powerful
relativistic jets.

\section*{Acknowledgments}
Based on observations obtained with \textit{XMM-Newton}, an ESA
science mission with instruments and contributions directly funded by
ESA Member States and the USA (NASA).  DRB acknowledges financial
support from the Natural Sciences and Engineering Research Council of
Canada. ACF thanks the Royal Society for support.


\bsp 

\label{lastpage}

\end{document}